\documentclass{SCGE}
\usepackage{graphicx}
\begin{document}

\begin{picture}(0,0){\rm
\put(0,-20){\makebox[160truemm][l]{\bf {\sanhao\raisebox{2pt}{.}}
Article  {\sanhao\raisebox{1.5pt}{.}}}}}
\put(0,-34){\jiuwuhao {\textcolor[rgb]{0.5,0.5,0.5}{\sf Progress of Projects Supported by NSFC
}}}
\end{picture}

\def\bm{\boldsymbol}

\def\dl{\displaystyle}
\def\du{\end{document}}
\def\d{{\rm d}}
\def\e{{\rm e}}
\def\i{{\rm i}}

\def\pi{{\uppi}}

\Year{2015} %
\Month{January} %
\Vol{58} 
\No{1} 
\BeginPage{1} 
\AuthorMark{{\rm Tong H}}  
\AuthorMarkCite{{\rm Tong H}. } 
\DOI{10.1007/s11433-014-5625-8} 
\ArtNo{000000}

\title[Pulsar braking: magnetodipole vs. wind]{Pulsar braking: magnetodipole vs. wind}

\author[]{TONG Hao}{}

\address[]{Xinjiang Astronomical Observatory, Chinese Academy of Sciences, Urumqi, Xinjiang 830011, China; tonghao@xao.ac.cn}

\maketitle \vspace{-3.5mm}{\footnotesize\begin{center} Received October 26, 2014; accepted November 5, 2014
\end{center}}\vspace*{-5mm}

\begin{center}
\rule{16.5cm}{0.4pt}
\parbox{16.5cm}
{\begin{abstract} 
Pulsars are good clocks in the universe. One fundamental question is that why they are good clocks? This is related to the braking mechanism of pulsars. Nowadays pulsar timing is done with unprecedented accuracy. More pulsars have braking indices measured. 
The period derivative of intermittent pulsars and magnetars can vary by a factor of several. However, during pulsar studies, the magnetic dipole braking in vacuum is still often assumed.  It is shown that the fundamental assumption of magnetic dipole braking (vacuum condition) does not exist and it is not consistent with the observations. The physical torque must consider the presence of the pulsar magnetosphere. Among various efforts, the wind braking model can explain many observations of pulsars and magnetars in a unified way. It is also consistent with the up-to-date observations. It is time for a paradigm shift in pulsar studies: from magnetic dipole braking to wind braking. As one alternative to the magnetospheric model, the fallback disk model is also discussed. 
\end{abstract}}
\end{center}\vspace*{-0.6cm}

\begin{center}
\parbox{16.5cm}
{\bf\jiuhao magnetar, magnetic field, neutron star, pulsar, wind}
\end{center}

\begin{center}
{\PACS{\rm 23.40.-s, 23.40.Bw}}
\CITA    
\end{center}

\textwidth=178truemm \textheight=236truemm

\wuhao\vspace*{1.5mm}

\begin{multicols}{2}

\renewcommand{\baselinestretch}{1.08} \baselineskip 12.2pt\parindent=10.8pt

\renewcommand{\thefootnote}

\section{Introduction}

\begin{figure}[H]
\centering
\includegraphics[width=0.45\textwidth]{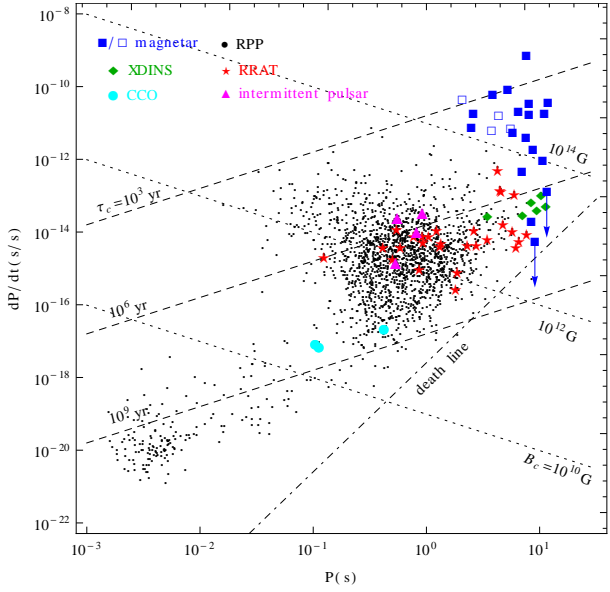}
\caption{Period and period derivative diagram of pulsars, including normal pulsars (black points), magnetars (blue squares, empty squares for radio loud magnetars), X-ray dim isolated neutron stars (green diamonds), central compact objects (light blue circles), rotating radio transients (red stars) and intermittent pulsars (magenta triangles). Updated from Figure 1 in \cite{TongWang2014}.}
\label{figPPdot}
\end{figure}

Pulsars are rotating magnetized neutron stars. They are good clocks in the universe. Because of their high precision timing, pulsars are good probes of interstellar medium and  magnetic field \cite{Han2013}, and good probes of gravitational wave background \cite{Lee2013}.
At the same time, pulsars have many other applications, e.g. pulsars as clocks etc. Among these applications, one fundamental question is that why pulsars are good clocks? This is related to the emission mechanism of pulsar multi-wave radiations and the spin down of pulsars.  
For the spin down mechanism of pulsars, one may ask, taking the Crab pulsar as an example, ``how to explain its period derivative, braking index etc?''  The period and period derivative of various pulsar-like objects are shown in figure \ref{figPPdot}. 

Pulsars must have magnetospheres (a system of particles surrounding the central neutron star). In the magnetosphere, there is particle acceleration. The radiation of these particles is responsible for the pulsar emissions. However, the magnetic dipole braking in vacuum is often assumed in pulsar timing studies, which is one dilemma in current pulsar studies. In the following, it is shown that the fundamental assumption of magnetic dipole braking (vacuum condition) does not exist, and it is not consistent with the pulsar observations. A physical spin down mechanism must be based on the existence of a magnetosphere\cite{Goldreich1969}. One candidate at present is the wind braking model \cite{Xu2001,Tong2013}. 

Section two is about the early understanding of pulsars. Observational progresses are provided in section three, including the braking index of pulsars, the spin down behaviors of intermittent pulsars and magnetars. Theoretical progresses are given in section four, including pseudo-magnetic dipole braking (various modifications of magnetic dipole braking), wind braking of pulsars, wind braking of magnetars and some discussions of the fallback disk model. Summary and prospects are presented in section five. The following materials (especially section two) are influenced by  three textbooks \cite{Xu2006,Shapiro1983,Lyne2012}. During the preparation of this paper, we read the work of \cite{Beskin2010}(mainly chapter two there). One aim of this paper and \cite{Beskin2010} is the same: trying to prove that the magnetic dipole braking in vacuum is incomplete (even irrelevant) for the study of pulsar spin down.

\section{A bite of pulsar early researches}

When Hewish et al.\cite{Hewish1968} first reported their discovery of the pulsating radio source (i.e. pulsar), the oscillations
of white dwarfs or neutron stars were thought to be responsible for the observations.
It was Gold\cite{Gold1968} who first pointed out the rotating neutron stars as the origin of pulsars
(the rotating neutron star scenario was not welcomed by the pulsar community when it was first proposed). 
If the radio emission
comes from the rotational energy of the neutron star, then two natural predictions will be made\cite{Gold1968}:
(1) pulsars will slow down gradually; (2) more rapidly rotating pulsars are expected (the pulsar period at that time
ranges from $0.25\,\rm s$ to $1.33 \,\rm s$). Later discovery of the Vela pulsar (with period of $89\,\rm ms$),
the Crab pulsar (with period $33\,\rm ms$), especially the slowdown rate of the Crab pulsar
(about one part in 2400 per year) has confirmed the rotating neutron star origin of pulsars \cite{Gold1969}.

With the period and period derivative of the Crab pulsar, the central star's rotational energy is
\begin{eqnarray}
 E_{\rm rot} &=& \frac{1}{2} I \Omega^2 = 2\pi^2 I \frac{1}{P^2} \\
 &=&2\times 10^{46} \frac{1}{P^2} \rm \, erg \approx 2\times 10^{49} \rm \, erg,
\end{eqnarray}
where $I$ is the neutron star's moment of inertia (which is order of $10^{45} \rm\, g \,cm^2$, for a star mass
$1-2 \rm \,M_{\odot}$ and radius about $10\rm\,km$), $\Omega$ is the angular velocity, and $P$ is the rotation period.
Following the tradition of theoretical astrophysics, CGS units are used in this paper.
The corresponding rotational energy loss rate is
\begin{eqnarray}
 |\dot{E}_{\rm rot}|& =& - \dot{E}_{\rm rot} = 4\pi^2 I \frac{\dot{P}}{P^3} \label{Edot}\\
 &=& 4\times 10^{46} \frac{\dot{P}}{P^3} \rm \, erg \,s^{-1} \approx 5\times 10^{38} \rm \, erg \,s^{-1}.
\end{eqnarray}
It is much higher than the radio luminosity. The energy output meets the energy budget required by the Crab nebula.
Therefore, from the beginning of pulsar astronomy, it is already known that the radiation energy is only a small fraction of the
rotational energy loss rate (especially for the radio emissions). Most of the rotational energy is taken away by
the particles and electromagnetic waves. This mixture of particles and waves is named as particle wind (or wind) in the following.
This particle wind may be visible in the form of pulsar wind nebulae\cite{Gaensler2006}.

\subsection{Rotating magnetic dipole in vacuum}

No mathematical formula was presented in Gold's theoretical paper (this is Gold's style).
The magnetic dipole radiation formula was employed by Pacini to consider the extraction of
rotational energy and slowdown of neutron stars before and after the discovery of pulsars.
In 1967, just before the discovery of pulsars, Pacini \cite{Pacini1967} proposed the idea of extracting rotational
energy of neutron stars by magnetic dipole radiation. This may provide a continuous energy supply
required by the Crab nebula observations. For a rotating magnetic dipole in vacuum, the radiation
frequency is equal to the angular speed of the neutron star. The corresponding radiation luminosity
is
\begin{equation}\label{Edip}
 \dot{E}_{\rm d} = \frac{2\mu^2 \Omega^4}{3c^3} \sin^2\alpha,
\end{equation}
where $\mu$ is dipole magnetic moment, $\alpha$ is the angle between the rotational axis and the magnetic axis (i.e. the inclination angle), and $c$ is the speed of light.
The dipole magnetic moment is related to the equatorial surface magnetic field $B_{\rm s}$ and polar surface magnetic field $B_{\rm p}$ as: $\mu =B_{\rm s} R^3 =1/2 B_{\rm p} R^3$ ($R$ is the neutron star radius). The magnetic field at the magnetic pole is two times as that at the magnetic equator $B_{\rm p} =2B_{\rm s}$ (for a dipole geometry). 
Equation (\ref{Edip}) can be rewritten in terms of magnetic field strength
\begin{equation}\label{Edip_in_B}
 \dot{E}_{\rm d} = \frac{2B_{\rm s}^2 R^6 \Omega^4}{3 c^3} \sin^2\alpha
                   =\frac{B_{\rm p}^2 R^6 \Omega^4}{6 c^3} \sin^2\alpha.
\end{equation}
The dipole magnetic moment is about $10^{30} \rm \,G \,cm^3$ for a magnetic field of $10^{12} \rm\, G$. The corresponding dipole radiation luminosity
is about $10^{36}$-$10^{40}\rm\,erg \,s^{-1}$ for an angular speed of $10^2$-$10^3 \rm\,rad \,s^{-1}$.

After the discovery of pulsars, \cite{Pacini1968} considered the magnetic dipole braking of pulsars further.
By equating equation (\ref{Edot}) and (\ref{Edip}), the period evolution of the neutron star is
\begin{equation}\label{Pdot_evolution}
 \dot{P} = \frac{8\pi^2 \mu^2 \sin^2\alpha}{3 I c^3} P^{-1}.
\end{equation}
If the timing parameters are known, the surface magnetic field
can be obtained from equation (\ref{Pdot_evolution})
\begin{equation}
 B_{\rm s} \sin\alpha = 3.2\times 10^{19} \sqrt{P \dot{P}} \,\rm G.
\end{equation}
Assuming an inclination angle of 90 degrees, the commonly employed expression for
the characteristic magnetic field of pulsars is obtained
\begin{equation}
 B_{\rm c}({\rm equator}) = 3.2\times 10^{19} \sqrt{P \dot{P}}\,\rm G.
\end{equation}
Therefore, the characteristic magnetic field is the equatorial magnetic field by equaling all the
braking torque to an orthogonal rotator in vacuum. Furthermore, for the study of pulsar emissions,
the magnetic field strength at the magnetic pole is more relevant. It is two times larger than
the equatorial surface magnetic field
\begin{equation}
B_{\rm c} ({\rm pole}) = 6.4\times 10^{19} \sqrt{P \dot{P}}\,\rm G. 
\end{equation}

The rotational evolution of a pulsar can be obtained by
integrating equation (\ref{Pdot_evolution})
\begin{eqnarray}
 P(t) &=& \sqrt{P_0^2 +\frac{16 \pi^2 \mu^2 \sin^2\alpha}{3 I c^3} (t-t_0)} \label{P_evolution_analytical}\\
 \label{P_evolution}
 &=& \sqrt{P_0^2 + 6.16 \times 10^{-8} B_{\rm s,12}^2 \sin^2\alpha (t-t_0)},
\end{eqnarray}
where $P_0$ is the initial rotational period at time $t_0$, and $B_{\rm s, 12}$ is the surface equatorial magnetic
field in units of $10^{12} \,\rm G$. Equation (\ref{P_evolution})
is obtained for $(t-t_0)$ in units of years. For a given set of parameters ($P_0, \ t_0, \ B_{\rm s}, \ \alpha$),
when $t$ is small, the rotational period $P(t)$ changes very little $P(t) \approx P_0$.
During the late time, when $t$ is large, the period increases with the time as $P(t) \propto t^{1/2}$.
The evolution of period derivative, rotational energy, and rotational energy loss rate etc can all
be obtained directly from equation (\ref{P_evolution}). From equation (\ref{P_evolution_analytical}),
the age of a pulsar is (using equation(\ref{Pdot_evolution}))
\begin{equation}
 T\equiv t-t_0 = \tau_{\rm c} \left(1 - \left(\frac{P_0}{P} \right)^2 \right),
\end{equation}
where 
\begin{equation}
\tau_{\rm c} \equiv \frac{P}{2\dot{P}}
\end{equation}
is defined as the characteristic age.
If the magnetic dipole braking in vacuum is valid, and the initial period is much
less than the present period $P_0 \ll P$, then $T=\tau_{\rm c}$ \cite{Gunn1969}.
On the other hand,
if the initial period is close to the present period $P_0 \approx P$, then
$T \ll \tau_{\rm c}$, i.e. the characteristic age may be much larger than the pulsar's true age.
The central compact objects may correspond to this case\cite{Gotthelf2013}.

The magnetic dipole braking can be further developed by considering the presence of
gravitational wave radiations and free precession 
due to deformation of the neutron star by its magnetic field,
the decay of the magnetic field, and radiation reaction on the inclination angle
\cite{Ostriker1969,Davis1970}.
All these three effects will result in a braking index larger than three
(see equation(\ref{braking_index})).
However, the magnetic dipole braking assumes a rotating dipole in vacuum.
This may make it not only incomplete but also irrelevant for the braking of
real pulsars \cite{Davis1970}. 

\subsection{The presence of pulsar magnetospheres}

There are many pioneering works since \cite{Goldreich1969}.
The scale height of the neutron star atmosphere is about $1\,\rm cm$.
The gravitational binding energy of a proton (or electron) on the neutron star surface is about
$100\,\rm MeV$ (or $0.1\,\rm MeV$). However, the electromagnetic force
will dominate over the gravitational force. Dimensional analysis shows that \cite{Shapiro1983}, 
the ratio of electric force to the gravitational force
for a proton is $\frac{e B_{\rm p} \Omega R/c}{G M m/R^2} \sim 10^9$, where
$e$ is the absolute value of electron charge, $G$ is the gravitational constant,
$M$ is the neutron star mass, and $m$ is the proton mass.
Therefore, a magnetosphere must be created around the central neutron star \cite{Goldreich1969,Beskin2010}.
The magnetosphere tends to corotate with the central neutron star.
However, the rotational velocity cannot exceed the speed of light.
The light cylinder radius is defined as where the rotational velocity equals
the speed of light
\begin{equation}\label{light_cylinder}
 R_{\rm lc} = \frac{c}{\Omega} =\frac{P c}{2\pi} =4.8\times 10^9 P\,\rm cm.
\end{equation}
For the magnetic field lines closed within the light
cylinder, they will corotate with the neutron star (``closed field line regions''). 
In the closed field line regions, in a steady state, the Lorentz force vanishes \cite{Shapiro1983}(which
follows the treatment of \cite{Goldreich1969})
\begin{equation}\label{electric_field}
 \bm{E} + \frac{\bm{\Omega} \times \bm{r}}{c} \times \bm{B} =0,
\end{equation}
where $\bm{E}$ is the electric field. The corresponding space charge density is
(named as the Goldreich-Julian charge density) 
\begin{equation}
 \rho_{e} = \frac{1}{4\pi} \bm{\bigtriangledown \cdot E} \approx -\frac{\bm{\Omega \cdot B}}{2\pi c}.
\end{equation}
The corresponding particle density is
\begin{equation}
 n_{e} =7\times 10^{-2} B_{z} P^{-1} \,\rm cm^{-3},
\end{equation}
where $B_{z}$ is magnetic field component parallel to the rotational axis. The Goldreich-Julian density is 
the charge density, which is the net value between positive and negative charges. 
The particle density of positive and negative charges can be much higher. 

However, this electric field (eq.(\ref{electric_field})) is perpendicular to the magnetic field
$\bm{E \cdot B} =0$. In the corotating frame (an inertia frame corotating with the neutron star
at some instantaneous time), the electric field is $\bm{E^{\prime}} = \bm{E}+ \frac{\bm{v}}{c} \bm{\times B} =0$.
In the closed field line regions, particles move the along field lines. The perpendicular electric field
has no effect on particle acceleration. The electric field responsible for the particle acceleration is
the parallel component $\bm{E \cdot B} \neq 0$ (which is a Lorentz invariant). In the open field line
regions, when the charge density deviates from the Goldreich-Julian density, it will cause the appearance
of a parallel electric field. Separating the total electric field in the open field line regions as
$\bm{E}= \bm{E}_{\rm acc}+ \bm{E}_{\rm GJ}$, where $\bm{\bigtriangledown \cdot E}_{\rm GJ} = 4\pi \rho_{\rm GJ}$,
the accelerating electric field satisfies the equation
\begin{equation}\label{possion}
 \bm{\bigtriangledown \cdot E}_{\rm acc} =4\pi (\rho-\rho_{\rm GJ}).
\end{equation}
Maybe, the most simple case would be a gap, where there is no net charge, i.e. $\rho=0$.
Near the neutron star surface, when the acceleration region is small, eq.(\ref{possion})
can be rewritten in a one-dimensional form
\begin{equation}
\frac{\d E_{\parallel}}{\d s} = 4\pi (\rho-\rho_{\rm GJ}),
\end{equation}
where $E_{\parallel}$ is the electric field component parallel to the magnetic field line,
$s$ is the distance measured along the magnetic field line, and in the gap $\rho=0$.
The gap bottom (the neutron star surface)
corresponds to $s=0$. Denote the gap height as $s=H$. If the electric field at the top
of the gap is defined as $E_{\parallel}(s=H)=0$, then the analytical solution for the electric field
can be obtained for the aligned rotator (the magnetic axis and the rotational axis are parallel
to each other)\cite{Xu2006} (adopting a similar procedure as\cite{Ruderman1975})
\begin{equation}
 E_{\parallel} = \frac{2\Omega B_{\rm p}}{c} (s-H).
\end{equation}
The potential difference between the top and bottom of the gap is
\begin{equation}
 \Delta \phi = \frac{\Omega B_{\rm p}}{c} H^2.
\end{equation}
For a pulsar with a rotational period of one second, $B_{\rm p} =10^{12} \,\rm G$, and a gap
height of $H=10^4\,\rm cm$, the potential difference is about $\Delta \phi =6\times 10^{12} \,\rm V$.
The gap height is determined by the complicated physical processes in the magnetosphere.
It is related to the specific particle acceleration model.
Various physically motivated acceleration models are proposed since \cite{Ruderman1975}.

The magnetospheric torque can be estimated in several ways. The electromagnetic field far away from
the neutron star may be of a transverse wave form\cite{Goldreich1969,Shapiro1983}.
The local electric field will be the same as the magnetic field $E\sim B$ (in CGS units).
The corresponding Poynting flux is $S\sim \frac{c}{4\pi} B^2$. At the light cylinder,
the energy loss rate is
\begin{equation}\label{Edot_magnetosphere_1}
 \dot{E} \sim S 4\pi R_{\rm lc}^2 = \frac{B_{\rm p}^2 R^6 \Omega^4}{c^3}.
\end{equation}
The dipole magnetic field decreases with the radius as $B(r)\approx B_{\rm p} (R/r)^3$.
This relation is used in the deduction of the above equation.
Equation (\ref{Edot_magnetosphere_1}) has the same form as the vacuum dipole case
(eq.(\ref{Edip_in_B})).

Another way of calculating the magnetospheric torque is by counting the energy carried away
by each outflowing particle. It is determined by the geometry and particle acceleration in the
open field line regions.
The footprint of all the open field lines makes up a region, dubbed as the polar cap.
For a rotating dipole, the field line equation is $r=r_{\rm d} \sin^2\theta$,
where $r_{\rm d}$ is the maximum radial extension. When the maximum
radial extension equals the light cylinder radius, it is defined as the last closed field
line (or the last open field line). The corresponding colatitude at the neutron star surface
is the polar cap angle
\begin{equation}
 \theta_{\rm pc} =\sin^{-1} \sqrt{\frac{R}{R_{\rm lc}}} \approx \sqrt{\frac{R}{R_{\rm lc}}}
 =1.4 \times 10^{-2} \, P^{-1/2},
\end{equation}
where the neutron star radius is taken as $R=10^6 \,\rm cm$.
The polar cap radius is $R_{\rm pc} = R\theta_{\rm pc} \approx 1.4\times 10^4 P^{-1/2} \,\rm cm$.
For the electric field satisfying eq.(\ref{electric_field}), the corresponding potential
drop between the polar cap edge and the magnetic pole is (it is actually the maximum acceleration
potential for an aligned rotating dipole \cite{Ruderman1975})
\begin{equation}
 \Delta \Phi_{\rm max} =\frac{R^2 \Omega B_{\rm p}}{2c} \theta_{\rm pc}^2
                       = 6.6\times 10^{12} \frac{B_{\rm p,12}}{P^2} \,\rm V,
\end{equation}
where $B_{\rm p,12}$ is the magnetic field in units of $10^{12} \,\rm G$ (in later sections, it will be simply written as 
$B_{\rm 12}$). The potential
is expressed in units of volts. During numerical calculations,
it should be converted back to CGS units.
The particle flux in the polar cap region can be written as $\rho \, c= \kappa\, \rho_{\rm GJ} \, c$,
where the charge density is parametrized as $\kappa$ times the Goldreich-Julian density.
When $\kappa=1$ and the acceleration potential equals the maximum acceleration potential,
the corresponding energy loss rate due to the particle outflow is
\begin{equation}\label{Edot_particle}
 \dot{E} =2 \pi R_{\rm pc}^2 \rho_{\rm GJ} c \Delta\Phi_{\rm max}
         =\frac{B_{\rm p}^2 R^6 \Omega^4}{2c^3}.
\end{equation}
It is also similar to the vacuum dipole case (eq.(\ref{Edip_in_B})).

In the magnetosphere, particles must attain some critical acceleration potential
(about $10^{12}$-$10^{13}\,\rm V$) in order to generate secondary particles 
\cite{Sturrock1971,Ruderman1975}.
For a constant acceleration potential $\Delta\phi=$constant, the corresponding energy loss rate is
(after some deduction similar to eq.(\ref{Edot_particle}))
\begin{equation}\label{Edot_constant_phi}
 \dot{E} = \frac{B_{\rm p} R^3 \Omega^2}{c} \Delta\phi.
\end{equation}
The slow down of a pulsar can be expressed in a power law form \cite{Shapiro1983}
\begin{equation}
 \dot{\Omega} \propto -\Omega^n,
\end{equation}
where $n$ is called the braking index, the minus sign corresponds to the slow down of the pulsar.
Observationally, the braking index of a pulsar is defined as \cite{Lyne2012}
\begin{equation}\label{braking_index}
n=\frac{\Omega \ddot{\Omega}}{\dot{\Omega}^2} = \frac{\nu \ddot{\nu}}{\dot{\nu}^2}= 2- \frac{P \ddot{P}}{\dot{P}^2}. 
\end{equation}
In pulsar studies, the angular velocity $\Omega=2\pi/P=2\pi \,\nu$, pulse frequency $\nu=1/P$, and the rotational period $P$ are frequently used in different conditions. Pulsar researchers should have no difficulty in transforming one expression to another. 
For the dipole braking case, the braking index is $n=3$. The braking index is $n=1$ for the
magnetospheric torque in the case of constant acceleration potential (eq.(\ref{Edot_constant_phi})).
Therefore,
the magnetic dipole braking and the magnetospheric torque are 
different on the second order effect, i.e. the braking index.
This defines the condition when the magnetic dipole braking approximation can be used. 

\section{Observational progresses}

\subsection{Braking index measurement of pulsars}

For a rotating dipole in vacuum, the theoretical braking index is exactly three, if the magnetic field,
moment of inertial, and inclination angle are constant. 
\cite{Groth1975} correctly measured the braking index of the Crab pulsar : $n=2.515\pm0.005$. It is different from 
the prediction of magnetic dipole braking. This strengthens the belief that the rotating dipole 
in vacuum is incomplete in modelling the spin down of pulsars. 
A second source had its braking reported in \cite{Manchester1985}. For PSR B1509-58, its braking index was 
$n=2.83\pm0.03$. It is similar to the Crab pulsar. In the presence of 
a particle wind, a smaller braking index $n=1$ is expected \cite{Michel1969} (and previous discussions, eq.(\ref{Edot_constant_phi})). In the presence of a pulsar magnetosphere,
there may naturally be some kind of particle outflow. These two early  
observations also indicate that it is hard to measure the braking indices of pulsars. It requires the determination 
of period second derivatives. The period second derivatives of most of the sources are dominated by ``noise" processes
\cite{Hobbs2010}. 

There are eight pulsars with braking index measured to the end of 2014 \cite{Lyne2015}. The braking index now ranges from
$n=0.9\pm 0.2$ for PSR J1734-3333 \cite{Espinoza2011} to $n=2.839\pm0.003$ for PSR B1509$-$58\cite{Livingstone2005}. 
In addition to the increase of the number of sources, the current braking index observations also show something new.     
\begin{enumerate}
  \item The smallest braking index is for PSR J1734-3333: $n=0.9\pm 0.2$ \cite{Espinoza2011}. 
  It is consistent with a wind dominated magnetospheric torque \cite{Kou2015}. 
  \item There are hints for the evolution of pulsar braking index \cite{Espinoza2013}. The braking index may evolve 
  from about three for young sources to about one for old sources. The evolution of pulsar braking index should be 
  considered in theoretical models. 
  \item The braking index is smaller during recovery from glitches. For PSR J1846-0258, its braking index decreased from $n=2.65\pm 0.01$
  to $n=2.16\pm0.13$ after glitches \cite{Livingstone2011}. The braking index of the Crab pulsar is about 
  $n=2.3$ during a glitch active epsiode\cite{Lyne2015}.  This may be viewed as evidence for glitch induced magnetospheric 
  activities in normal pulsars \cite{Wang2012}. 
  \end{enumerate}
More observations are required to confirm these interesting aspects. 

\subsection{Spin down of intermittent pulsars}

In the early age of pulsar studies,
some pulsars were found to stop working for several periods, i.e. nulling \cite{Backer1970}. Intermittent pulsars may be viewed as 
a special kind of nulling pulsars. The first intermittent pulsar PSR B1931+24 was discovered long ago. 
 \cite{Kramer2006} reported that PSR B1931+24
showed two distinct states. In the ``on'' state, the pulsar can be detected and viewed as an ordinary pulsar. In the ``off'' state,
the pulsar is not detected (various upper limits for the radio flux are reported). More intriguingly, 
the pulsar spin down rate is larger in the on state than that in the off state. The ratio of spin down rate between the on and off state 
is $\dot{P}_{\rm on}/\dot{P}_{\rm off}=1.5$ \cite{Kramer2006}. This means that there may be some additional particle outflows during the on state. This particle outflow is responsible for both the radiation of radio emissions and larger spin down rate (because these outflowing particles will take away additional angular momentum of the central neutron  star \cite{Li2014}). 
The intermittent pulsars can have accurate
determination of spin down rate for both the on and off state because their time scale of on and off state is very large. 
For PSR B1931+24, its on state lasts about 5 days. The off state lasts about 30 days. The duty cycle of the on state is 
about $20\%$ (the fractional time when the pulsar is in the on state). 

Up to the end of 2014, three intermittent pulsars are reported \cite{Kramer2006,Camilo2012,Lorimer2012}. 
The latter two intermittent pulsars can be in the off state  for about one year. 
The spin down ratio ranges from
$\dot{P}_{\rm on}/\dot{P}_{\rm off}=1.5$ (for PSR B1931+24 \cite{Kramer2006}) to  
$\dot{P}_{\rm on}/\dot{P}_{\rm off}=2.5$ (for PSR J1841-0500 \cite{Camilo2012}).  Timing and radiation are two major aspects of pulsars. Intermittent pulsars clearly show the correlation between timing and radiative behaviors. This kind of correlation may exist in all pulsars (only different by the magnitude of variations)\cite{Lyne2010}. 
At the same time, efforts are made toward finding similar spin down behaviours in nulling pulsars (e.g.
a larger spin down rate when the pulsar is on \cite{Young2015}).  
Furthermore, rotating radio transients \cite{McLaughlin2006} and magnetar pulsed radio emission \cite{Camilo2006} were also discovered. Along with intermittent pulsars, they all belong to the increasing diversity of transient radio pulsars. More transient sources may await to be discovered in future surveys (good news for future survey project, e.g. \cite{Karako2015}). Future more accurate timing and multi-wave observations may unveil the nature of these transient radio pulsars (also helpful for ``persistent'' radio pulsars). 

\subsection{Timing behaviors of magnetars}

More than twenty magnetar candidates are discovered \cite{Olausen2014} (http://www.physics.mcgill.ca/$\sim$pulsar/magnetar/main.html). They form a separate class of pulsars (neutron stars powered by their magnetic energy \cite{Duncan1992}). Magnetars manifest themselves as anomalous X-ray pulsars and soft gamma-ray repeaters  \cite{Katz1982,TongXu2014}. Though their number is small, each source has its own peculiarities.
\begin{description}
\item[ Radio loud magnetars] These sources belong to the increasing number of transient  magnetars. 
The first transient magnetar XTE J1810$-$197 was reported in 2004 \cite{Ibrahim2004}. Later it was reported to show 
pulsed radio emissions \cite{Camilo2006}. This discovery of radio emitting magnetar bridged the gap between magnetars and normal radio pulsars (the majority of rotation-powered pulsars). It also provides new clues to the study of pulsar radio emissions (spectra, transient, and why radio emissions at all). Later multiwave observations found a decreasing spin down rate (by a factor of three; here the spin down rate refers to the period derivative) while the star's radio and X-ray luminosity keep decreasing \cite{Camilo2007}. Similar decreasing spin down rate was also found in another radio emitting magnetar PSR J1622$-$4950 \cite{Levin2012} when the star's radio and X-ray flux decrease with time \cite{Anderson2012}. A decreasing spin down rate during outburst is naturally expected in the wind braking model of magnetars \cite{Tong2013}. 

For the radio magnetar 1E 1547.0$-$5408, its spin down rate increases when its X-ray flux is decreasing with time \cite{Camilo2008}. Similar negative correlation was also found in the radio magnetar near the Galactic center SGR J1745$-$2900. Its period derivative may have increased by a factor of two while the star's X-ray luminosity is decreasing with time \cite{Kaspi2014}. 
This negative correlation between radiative and timing behavior is contrary to the most simple idea of a magnetospheric torque. 
It may due to geometrical reasons \cite{Tong2015a} or accumulation of magnetic energy in the magnetosphere (which may leads to outbursts in the future \cite{Tong2015b}). 

\item[Low magnetic field magnetars]  Traditionally, magnetars were thought to be young neutron stars with a strong magnetic field \cite{Kouveliotou1998}. The difference between the dipole field and the multipole field is not clearly distinguished.  This situation changed with the discovery of a magnetar with dipole magnetic field lower than $7.5\times 10^{12} \,\rm G$, i.e. a low magnetic field magnetar \cite{Rea2010}. The presence of a low magnetic field magnetar may be due to physical \cite{Turolla2011} or geometrical reasons \cite{Tong2012}. It clearly shows that a strong dipole field is not the essential ingredient of magnetars. In the magnetar model, there are various kinds of magnetic fields (core poloidal/toroidal field, crustal poloidal/toroidal field, magnetospheric poloidal/toroidal field, and local magnetic field domains etc). There are many parameters in the magnetar model instead of one. More low magnetic field magnetars were discovered later \cite{Rea2012,Zhou2014}.  

\item[Enhanced spin down rate and anti-glitch] Many magnetars show enhanced spin down rate during burst active episode (1E 2259$+$586\cite{Kaspi2003};  SGR 1806$-$20\cite{Woods2007}). Marginal evidence of net spin down of the central neutron star was found in SGR 1900$+$14 previously \cite{Woods1999}. Later clear evidence of net spin down was found in the continued monitoring of 
1E 2259$+$586 (dubbed as anti-glitch \cite{Archibald2013}). Similar net spin down was also observed in PSR J1846$-$0258 (which showed some magnetar activities at that time \cite{Livingstone2010}). Quasi-periodic torque variation was found in 1E 1048.1$-$5937
\cite{Archibald2015}. Anti-glitch may just be a period of enhanced spin down rate \cite{Tong2014}. The enhanced spin down rate may result from more activities in the magnetosphere. However, it is not clear at present why the magnetospheric activities can repeat quasi-periodically. 

\end{description}

\section{Theoretical progresses}

\subsection{Pseudo-magnetic dipole braking}

In the magnetic dipole braking model, the rotational energy loss rate is proportional to 
$\sin^2\alpha$. The inclination angle will evolve to a smaller value \cite{Davis1970}. 
The braking index will be larger than three considering the evolution of magnetic inclination angle (inconsistent with pulsar observations).
In order to explain the observations (e.g. the braking index), various modifications of the magnetic dipole braking model are considered. 

Generally, the spin down power law can be written as \cite{Kou2015} 
\begin{equation}
\dot{\Omega} = - \kappa(t) \Omega^3,
\end{equation}
where the coefficient $\kappa(t)$ may dependent on the magnetic field, inclination angle, angular velocity etc. 
The corresponding braking index will be 
\begin{equation}
n= 3- \frac{\tau_{\rm c}}{\tau_{\kappa}},
\end{equation}
where $\tau_{\rm c}$ is the characteristic age, and $\tau_{\kappa} = \kappa/2\dot{\kappa}$ is defined as the characteristic variation time scale of $\kappa(t)$. If $\dot{\kappa}>0$, then $\tau_{\kappa}>0$. This will result in $n<3$. This means that an increasing $\kappa$ will result in a braking index smaller than three. For the vacuum dipole case, $\kappa \propto \mu^2 \sin^2\alpha /I$. Therefore, an increasing magnetic dipole field strength \cite{Espinoza2011}, or an increasing inclination angle \cite{Lyne2013}, or a decreasing moment of inertia \cite{Yue2007} will result in a braking index smaller than three. Furthermore, the increase of magnetic field and increase of inclination angle are coupled \cite{Bhattacharya1991}.  

There are also other calculations done in the presence of a magnetosphere. At the same time, they can be understood  (at least qualitatively) in the vacuum dipole case. In the presence of a pulsar magnetosphere, the rotational energy loss rate is made up of two components (at least mathematically \cite{Xu2001,Contopoulos2006,Li2012}).
They are the magnetic dipole component (which is proportional to $\sin^2\alpha$) and the particle wind component (different models have different expressions for this component). As the pulsar approaches the death line, the particle wind component will gradually cease. For pulsars near the death line, their rotational energy loss rates are mainly dominated by the magnetic field perpendicular to the rotational axis, i.e. $B_{\rm p} \sin\alpha$. For a small inclination angle, the real surface magnetic field $B_{\rm p}$ can be much larger than the characteristic magnetic field, which is about $B_{\rm p} \sin\alpha$.  For the low magnetic field magnetar SGR 0418+5729, its characteristic magnetic field is only about $7.5\times 10^{12} \,\rm G$ \cite{Rea2010}. It already lies below the death line. For an inclination angle of 5 degrees, its surface dipole field can be as high $10^{14} \,\rm G$ \cite{Tong2012}. Therefore, it can be a normal magnetar instead of a low magnetic field magnetar if its inclination angle is small. 

Magnetohydrodynamical simulations of pulsar magnetosphere found a rotational energy loss rate proportional to $(1+\sin^2\alpha)$ \cite{Spitkovsky2006}. The term which depends on the inclination angle is the same as the magnetic dipole braking case, proportional to $\sin^2\alpha$. The evolution of inclination angle tends to reduce the rotational energy loss rate \cite{Philippov2014}. Then the inclination angle will also evolve to smaller value in the magnetohydrodynamical simulation. The consequent braking index will also be larger than three.  

The above two cases are just two examples. Along with various modifications of magnetic dipole braking, they may be dubbed as ``pseudo-magnetic dipole braking''. In the literature, there are many theoretical and observational papers where the calculation or discussion are based on the magnetic dipole braking assumption. It may give the reader the illusion that the magnetic dipole braking model is well established or at least not too bad. In reality, it is not the case. 
The magnetic dipole braking model may be viewed as a pedagogical model, especially for newcomers in pulsar astronomy. 

\subsection{Wind braking of pulsars}

Pulsars are oblique rotators in general. For a finite inclination angle, the magnetic moment can be decomposed into the perpendicular component (perpendicular to the rotational axis) and the parallel component. The rotational energy loss rate  due to the perpendicular magnetic moment may be approximated by the magnetic dipole radiation (see eq.(\ref{Edip}), which may actually be some kind of particle outflow) 
\begin{equation*}
\dot{E}_{\rm d} = \frac{2\mu^2 \Omega^4}{3c^3} \sin^2\alpha.
\end{equation*}
The parallel component may be responsible for the particle acceleration \cite{Ruderman1975}. The corresponding rotational energy loss due to the particle outflow is \cite{Xu2001}
\begin{equation}
\dot{E}_{\rm p} = 2\pi r_{\rm p}^2 c \rho \Delta \phi = \frac{2\mu^2 \Omega^4}{3 c^3} 3 \frac{\Delta \phi}{\Delta \Phi},
\end{equation}
where $\Delta \phi$ is the particle acceleration potential (depending on the specific acceleration model), and $\Delta \Phi$ is the maximum acceleration potential. The total rotational energy loss rate may be the combination of the perpendicular component and the parallel component \cite{Xu2001}:
\begin{equation}\label{Edot_simple}
\dot{E} = \frac{2\mu^2 \Omega^4}{3 c^3} \left( \sin^2\alpha + 3\frac{\Delta \phi}{\Delta \Phi} \cos^2\alpha  \right)
= \frac{2\mu^2 \Omega^4}{3 c^3} \eta, 
\end{equation}
where $\eta =\sin^2\alpha + 3\frac{\Delta \phi}{\Delta \Phi} \cos^2\alpha$. The magnetic dipole braking in vacuum corresponds to 
$\eta =\sin^2\theta$. The angular factor $\cos^2\alpha$ appears by considering that  the parallel component is mainly responsible for the particle acceleration. It is possible that the factor $\cos^2\alpha$ does not appear \cite{Li2014}. However, the difference is only quantitative. Equation (\ref{Edot_simple}) is obtained by assuming that the particle density equals the Goldreich-Julian charge density and the effect of pulsar death is not considered. For the polar gap model \cite{Ruderman1975}, the corresponding expression for $\eta$ is: $\eta= \sin^2\alpha+ 4.96\times 10^2 B_{12}^{-8/7} \Omega^{-15/7} \cos^2\alpha$ \cite{Li2014}. 
The corresponding expression can be obtained for each acceleration model. Therefore, precise pulsar timing observations may be used to distinguish between different particle acceleration models. There are also other versions of magnetospheric torque (similar to equation (\ref{Edot_simple}), \cite{Li2012,Beskin2010} and references therein). Every model has its own merits. They can be used to check each other. 

\begin{figure}[H]
\centering
\includegraphics[width=0.45\textwidth]{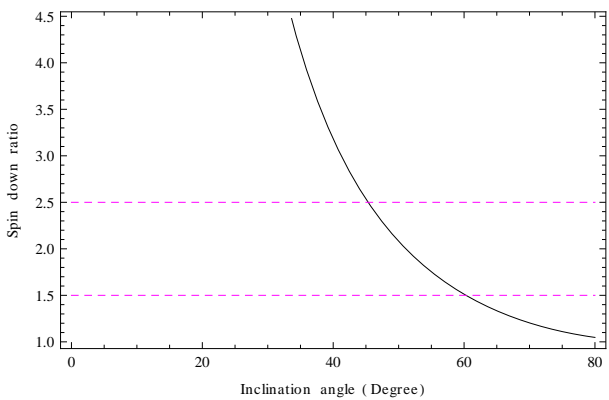}
\caption{Ratio of spin down rate for intermittent pulsars. The solid line is model calculation, for typical parameters of: 
$B=5\times 10^{12} \,\rm G$, $P=1\,\rm s$. The dashed line is the observational range of spin down ratio, from 1.5 to 2.5.
Adapted from figure 1 in \cite{Li2014}.}
\label{figspindownratio}
\end{figure}

\begin{figure}[H]
\centering
\includegraphics[width=0.45\textwidth]{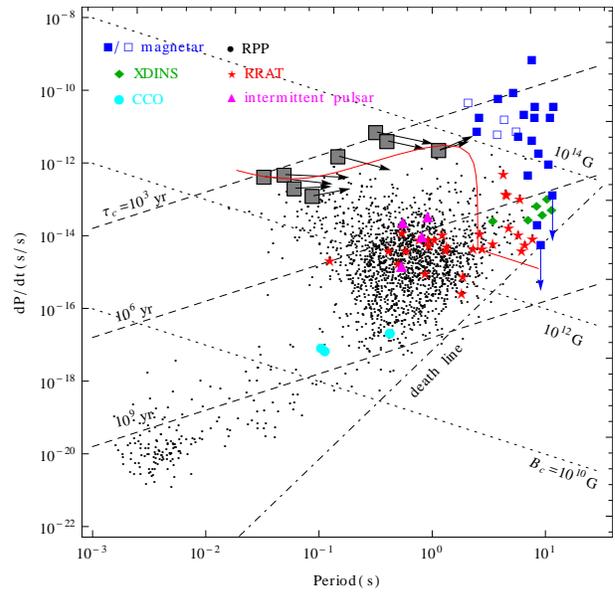}
\caption{Rotational evolution of the Crab pulsar in the wind braking model (red solid line). The eight sources with braking index measured are shown as grey squares with arrows marking their evolution direction. See figure \ref{figPPdot} for explanations of various pulsar-like objects. Updated from figure 6 in \cite{Kou2015}.}
\label{figPPdotCrab}
\end{figure}

\begin{description}
\item[Intermittent pulsars] During the on state, 
the presence of some additional particles may be responsible for both the turn on of radio emission and the enhanced spin down rate. The magnetic dipole
braking may be employed to model the spin down during the off state. Then the ratio of spin down rate between the on and off state is \cite{Li2014}:
\begin{equation}
 r \equiv \frac{\dot{\Omega}_{\rm on}}{\dot{\Omega_{\rm off}}} = \frac{\eta}{\sin^2\alpha}
 = \frac{\sin^2\alpha + 3(\Delta \phi/\Delta \Phi) \cos^2\alpha}{\sin^2\alpha} \ge 1. 
\end{equation}
Therefore, in the pulsar wind model, the spin down rate of intermittent pulsars during the on state is always larger than that during the off state. For the vacuum gap acceleration model \cite{Ruderman1975}, the expression of $\eta$ is known. For typical parameters, 
the corresponding ratio of spin down rate is a function of inclination angle, see Figure \ref{figspindownratio}. 
The predicted range of inclination angle can be checked by future observations. At the same time, the braking index 
of intermittent pulsar can also be predicted. 

\item[The Crab pulsar and other sources] Equation (\ref{Edot_simple}) may be further improved by considering the effect of particle density 
and pulsar death. Such a more perfect model of wind braking of pulsars is available in \cite{Kou2015}. It can be applied to the the Crab pulsar which has the most detailed timing observations.  
Figure \ref{figPPdotCrab} shows the rotational evolution of the Crab pulsar in the wind braking model. 
Many things can be seen in Figure \ref{figPPdotCrab} \cite{Kou2015}: (1) The spin down of the Crab pulsar will evolve 
from magnetic dipole braking dominated case (with braking index close to three) to wind dominated case (with braking index about one). 
The observed value of braking index ($2.51$ for the Crab pulsar) is the combined effect of magnetic dipole braking and particle wind. 
For the whole pulsar population, their braking index can be naturally divided into two groups: one group has a braking index close to three; the other group has a braking index close to one. Pulsars in the second group will be older than those in the first group. 
(2) The Crab pulsar (and all other pulsars) will not evolve to the cluster of magnetars. When the pulsar is sufficiently old, it will enter into
the death valley. This is also true for the magnetar population. Specific calculation for the low magnetic field magnetar has already 
been done \cite{Tong2012}(using another version of wind braking). (3) When the particle wind is stronger than the normal value (e.g.
during glitches), the braking index will be smaller. This stronger particle will also contribute to some net spin down of the pulsar. 
This may explain the lower braking index of the Crab pulsar during glitches. 
It may be viewed as glitches induced magnetospheric activities in normal pulsars . 
(4) The wind braking model can also be applied to other sources, e.g. pulsars with braking index measured, and the magnetar population. 
\end{description}

\subsection{Wind braking of magnetars}

During the timing study of magnetars, the magnetic dipole braking in vacuum is also often employed. Furthermore, a pulsar is often claimed to be a magnetar when its characteristic magnetic field is higher than the quantum critical value (which is $4.4\times10^{13} \,\rm G$, when the electron cyclotron energy equals its rest mass energy). These claims do not consider the difference between magnetars and high magnetic field pulsars \cite{Ng2011}. Even if the putative star is indeed a magnetar, its characteristic magnetic field cannot be used as a proof of its magnetar nature. This point is clearly demonstrated by the discovery of the low magnetic field magnetar 
SGR 0418$+$5729 (with a characteristic magnetic field less than $7.5\times 10^{12} \,\rm G$ \cite{Rea2010}). Furthermore, there are other observations arguing against a strong dipole field in magnetars \cite{Tong2013}:
the supernova energies associated with magnetars are of normal value \cite{Vink2006}; the non-detection of magnetars by {\it Fermi}-LAT \cite{Tong2010,Tong2011}; and the variable period derivative of magnetars (by a factor of several or more; on days or months time scale) etc. The variable spin down rate may be viewed as the direct evidence of magnetospheric torques \cite{Thompson2002,Tong2013}. It is hard to image that the large scale dipole field can vary, especially on short time scale. While, the magnetospheric torque (e.g. a system of particle outflow) may change dramatically even on short time scale. The system of particles can change dramatically because they originated from the magnetic energy. The X-ray luminosities of magnetars can 
vary significantly because it is also of magnetic origin. Therefore, in the wind braking model of magnetars, the timing and radiative behaviors of magnetars are correlated. This is commonly observed in magnetars \cite{Archibald2015}(and references therein). 

The magnetic energy release may first be converted into a system of non-thermal particles. These particles are responsible for the non-thermal radiation of magnetars. If the particle wind luminosity is $L_{\rm p}$, then the rotational energy loss rate due to this particle wind is \cite{Tong2013}
\begin{equation}\label{Edotwind}
\dot{E}_{\rm w} = \dot{E}_{\rm d} \left(  \frac{L_{\rm p}}{\dot{E}_{\rm d}} \right)^{1/2},
\end{equation}
where $\dot{E}_{\rm d}$ is the rotational energy loss rate due to a rotating dipole in vacuum. 
Two extremes of equation (\ref{Edotwind}) are:
\begin{description}
\item[Normal pulsar case] The particle wind originates from the rotational energy $L_{\rm p} = -\dot{E}_{\rm rot}$. Then the rotational energy loss rate due to the particle wind is order of magnitude the same as the vacuum dipole case. The effect of particle wind will mainly be seen in high order effects, e.g. braking index \cite{Xu2001}. Only in special sources, e.g. intermittent pulsars, the first order effect of particle wind  can be seen \cite{Li2014}. 

\item[Magnetar case] The particle wind is from the the magnetic energy release, therefore it can be much higher than the rotational energy loss rate $L_{\rm p} \gg -\dot{E}_{\rm rot}$. The rotational energy loss rate of magnetars will be dominated by the particle wind. A varying particle wind luminosity will result in a varying spin down rate of the putative magnetar. A strong dipole field is not necessary in order to explain the timing observations of magnetars. In the wind braking scenario, magnetars are neutron stars with a strong multipole field. The strong multipole field is responsible for the persistent X-ray luminosity, bursts (including giant flares), super-Eddington luminosity during bursts, and variable spin down etc.  
\end{description}
Therefore, equation (\ref{Edotwind}) represents a unification of wind braking of pulsars and wind braking of magnetars. 
Many consequences result from the wind braking of magnetars \cite{Tong2013}: 
\begin{enumerate}
\item The corresponding dipole magnetic field in the case of wind braking can be much lower (e.g. ten times lower or more) than the characteristic magnetic field. The surface dipole field at the magnetic pole is:
\begin{equation}
B_{\rm p} = 4.0\times 10^{25} \frac{\dot{P}}{P} \, L_{\rm p, 35}^{-1/2} \,{\rm G} 
= 4.0\times 10^{13} \frac{\dot{P}/10^{-11}}{P/10\,\rm s}\, L_{\rm p, 35}^{-1/2} \,\rm G,
\end{equation}
where $L_{\rm p, 35}$ is the particle luminosity in units of $10^{35} \,\rm erg\, s^{-1}$. 

\item For magnetars with a low X-ray luminosity, they tend to have similar magnetospheres to those of normal pulsars. 
Therefore, low luminosity magnetars are more likely to have radio emissions.

\item There are two predictions of the wind braking model of magnetars. A magnetism-powered pulsar wind nebula is expected to surround the central magnetar. The braking index of magnetars will be smaller than three. These two predictions are both hard to varify observationally. However, it is not impossible (especially in the future).

\end{enumerate}

Later observations are consistent with the wind braking model of magnetars.  
\begin{description}
\item[ Swift J1822.3$-$1606] This is the second low magnetic field magnetar \cite{Rea2012}. However, different authors found different period derivatives. While the observers were discussing whether this is caused by timing noise, \cite{TongXu2013} pointed out that the period derivative of Swift J1822.3$-$1606 may be decreasing with time. According to equation (\ref{Edotwind}), for a short timescale of years, the star's period derivative is related with the particle wind luminosity as: $\dot{P} \propto L_{\rm}^{1/2}$. Therefore, a decreasing particle luminosity after outburst will naturally result in a decreasing period derivative. 
\cite{TongXu2013} predicted a long term period derivative of  $1.9\times 10^{-14}$ (last paragraph, section 2 there). Recent timing observations found a period derivative of $(2.1\pm0.2) \times 10^{-14}$ \cite{Scholz2014}. This is consistent with the wind braking prediction.

\item[SGR J1745$-$2900] This source lies near the Galactic centre and is radio loud \cite{Rea2013}. During subsequent X-ray observations, SGR J1745$-$2900 show an increasing period derivative (two times larger) 
when its X-ray luminosity keeps decreasing \cite{Kaspi2014}. This negative correlation between timing and radiation is hard to understand. Considering the polar cap geometry of the particle wind, the particle wind may have a finite polar cap opening angle $\theta_{\rm s}$. Then the corresponding period derivative will be proportional to $\dot{P} \propto \theta_{\rm s}^{-4/3}$. Therefore, a smaller polar cap angle will result in a larger period derivative. The change of polar cap opening angle will cause a negative correlation between X-ray luminosity and the spin down rate \cite{Tong2015a}. In the wind braking model, SGR J1745$-$2900 will have a maximum spin down rate. Up to date timing observations are consistent with this prediction \cite{CotiZelati2015,Tong2015b}. 
  
\item[Anti-glitch]  The magnetar 1E 2259$+$586 suffered a net spin down during an observational interval about two weeks. This net spin down is dubbed as anti-glitch \cite{Archibald2013}. Previously, such spin down event also occurred in SGR 1900$+$14 \cite{Woods1999}, and PSR J1846$-$0258 \cite{Livingstone2010}. An enhanced particle wind during the observational interval will take away some amount of additional angular momentum. Therefore, the so call ``anti-glitch'' may just be a period of enhanced spin down of the central neutron star \cite{Tong2014}. The X-ray flux is also higher than the quiescent level. This is consistent with the wind braking interpretation of anti-glitch. 

Furthermore, the enhanced particle wind will result in a decreasing/varying braking index (if the star's braking index can be observed). 
This kind of decreasing/varying braking index is indeed observed in the Crab pulsar \cite{Wang2012,Lyne2015,Kou2015}. The opposite case of enhanced particle wind would be a period with a weaker particle outflow. This may correspond to the spin down behavior of intermittent pulsars \cite{Li2014}. Therefore, the wind braking model provides a unified explanation of (1) rotational evolution of the Crab pulsar, (2) spin down behavior of intermittent pulsars, and (3) anti-glitch.  

\end{description}

\subsection{Fallback disk model as an alternative to the magnetospheric model}

The above discussions are mainly in the wind braking model (or magnetospheric model). For these peculiar timing behaviors 
of pulsars and magnetars, there is always one alternative: the fallback disk model. The fallback disk model is also commonly employed to explain various pulsar-like objects \cite{Xu2007,Tong2011b} (as an alternative to the magnetar model). After the supernova explosion, some of the ejected material may try to fallback onto the neutron star. In the presence of some angular momentum, a disk around the central neutron star may be formed, i.e. fallback disk\cite{Wang2006}(and references therein). The observed pulsar braking index can be explained considering the fallback disk torque \cite{Fu2013,Liu2014}. The intermittent pulsar and other transient radio pulsars may result from the accretion of matter onto the neutron star from a fallback disk \cite{Li2006}. However, quantitative calculations in the fallback disk model are still lacking (e.g. the spin down ratio of intermittent pulsars has not been calculated quantitatively). More theoretical works are needed in order to confront with future multiwave observations. 

Anomalous X-ray pulsars and soft gamma-ray repeaters are proposed to be neutron star/fallback disk systems in the early age of magnetar research \cite{Chatterjee2000,Alpar2001}. Later timing observations of magnetars can also be understood in the fallback disk model (``low magnetic field magnetar'' \cite{Alpar2011}; anti-glitch \cite{Katz2014} etc). The fallback disk model may account for the persistent X-ray emission and timing of anomalous X-ray pulsars and soft gamma-ray repeaters naturally (possibly more natural than the magnetar model). In order to explain the giant flares and the $10^4$ times super-Eddington luminosity in the pulsating tail of giant flares, strong multipole fields as that of the magnetar model are required\cite{Mereghetti2008}(and references therein). Considering that the central neutron star may be a quark star\cite{Xu2007}, the quark star/fallback disk model may explain both the persistent and burst properties of anomalous X-ray pulsars and soft gamma-ray repeaters\cite{Tong2011b}. Future observations (e.g. X-ray polarimetry \cite{Lu2013}) may more clearly show which is closer to the truth, the magnetar model or the fallback disk model.

\section{Summary and prospects}

After about 50 years of pulsar discovery, pulsars timing are done with unprecedented accuracy. More pulsars have breaking index measured, variation of period derivatives detected, timing noise characterised etc. The existence of intermittent pulsars directly shows the evidence of particle wind. Furthermore, the timing behaviors of magnetars are always variable. However, the magnetic dipole braking in vacuum is still often employed to explain the timing observations (not only by newcomers but also by experts). It is also well known that the fundamental assumption of dipole braking (vacuum condition) does not exist. Therefore, it is time for a paradigm shift in pulsar studies: from magnetic dipole braking to wind braking. The wind braking model is just one of the models for calculating the magnetospheric torque \cite{Xu2001,Tong2013}. The confrontation between the wind braking model and pulsar timing observations is started:
\begin{enumerate}
\item The braking of index of pulsars (generally $1<n<3$)  is the combined effect of magnetic dipole braking and particle wind. 

\item In the case of intermittent pulsars, the ratio of spin down rate between the on and off state is determined by the inclination angle. 

\item The effect of pulsar death is considered when modelling the long term rotational evolution of pulsars. 

\end{enumerate}
More things are to be done in the future concerning the wind braking of pulsars:
\begin{itemize}
\item The wind braking model can be applied to more sources, e.g. the other seven pulsars with braking index measured, and the magnetar population.

\item The evolution of inclination angle in the presence of a particle wind. The evolution of rotational period and the inclination angle are coupled. The result may be compared with the Crab pulsar observations \cite{Lyne2013}. 

\item The dependence of pulsar death line on the equation of state. The neutron star mass, radius and moment of inertia will affect the consequent dipole magnetic field, acceleration potential etc. The position of death line on the period and period-derivative diagram, and the rotational evolution will both depend on the pulsar equation of state. 

\item The fluctuation of the magnetosphere may account for the timing noise of pulsars (e.g. \cite{Liu2011}). More detailed calculations should be done concerning the random fluctuation of the magnetosphere. 

\item In the presence of a strong magnetic field, the neutron star will be somewhat deformed. This will introduce other torques, e.g. gravitational wave emissions. At the same time, the deformed neutron star will also precess (free or forced). During the life time of a pulsar, its magnetic field may also evolve, e.g. decay in the long run. Therefore, a thorough modelling should include all these factors (some factor may only important in special cases). 
\end{itemize}

At present, the wind braking model of magnetars has achieved the aspects as follows:
\begin{enumerate}
\item Unification of wind braking of pulsars and wind braking of magnetars. 

\item A changing luminosity may explain the timing behavior of Swift J1822.3$-$1606. A changing geometry may correspond to the negative correlation between timing and X-ray luminosity in SGR J1745$-$2900. These are just two extreme cases. In reality, many factors can come into play. This may correspond to the various spin down behaviors of magnetars. 

\item The anti-glitch may be a period of enhanced spin down. 

\end{enumerate}
The wind braking model of magnetars may be further developed in the following aspects:
\begin{itemize}
\item The particle luminosity is the main parameter in the wind braking model. At present, it is either assumed as $10^{35} \,\rm erg \,s^{-1}$ (or other constant values) or $L_{\rm p} =L_{\rm x}$ (which varies with time as that of the X-ray luminosity). In the future, the particle luminosity may be calculated for a given state of magnetosphere, e.g. the present attempt of \cite{Beloborodov2009}. Its evolution with time can also be calculated. This will enable the wind braking model to be applied to more sources. 

\item The particle wind contributes both the radiation and braking torque of the magnetar. The present wind braking model mainly concentrate on the braking torque. The radiation of these particles should also be explored, e.g. radio, soft/hard X-ray. 

\item The pulsar wind nebulae is due to the particle outflow in normal pulsars. In the case of magnetars, there may also be a wind nebula. This wind nebula may even powered by the magnetic energy of the magnetar. Both observation and theory are in need in this direction. 
\end{itemize}

The wind braking model is among the efforts to unify different pulsar-like objects. 
The fallback disk model may provide an alternative to the magnetospheric model. 
 Accreting neutron stars form a distinct class of pulsars. Works for one special source  are like blind men grabbing a piece of the elephant, i.e. the pulsar. The ultimate understanding of pulsars must involve a global view of all kinds of pulsars.

\vspace*{2mm} \Acknowledgements{\bahao The author would like to thank R. X. Xu for encouragement; F. F. Kou, L. Lin, and R. Yuen for reading the manuscript; and the Referee for comments about the fallback disk model. H.Tong is supported by the Xinjiang Bairen project, West Light Foundation of CAS (LHXZ201201), Qing Cu Hui of CAS, and 973 Program (2015CB857100).}

\end{multicols}

\end{document}